# Big Data Analytics on Traditional HPC Infrastructure Using Two-Level Storage

Pengfei Xuan
School of Computing
Clemson University
Clemson, SC, 29634
USA
pxuan@g.clemson.edu

Jeffrey Denton
Omnibond Systems, LLC
81 Technology Drive
Anderson, SC, 29625
USA
denton@omnibond.com

Pradip K Srimani, Rong Ge, Feng Luo
School of Computing
Clemson University
Clemson, SC, 29634
USA
{psriman, rge, luofeng}@clemson.edu

## ABSTRACT
Data-intensive computing has become one of the major workloads on traditional high-performance computing (HPC) clusters. Currently, deploying data-intensive computing software framework on HPC clusters still faces performance and scalability issues. In this paper, we develop a new two-level storage system by integrating Tachyon, an in-memory file system with OrangeFS, a parallel file system. We model the I/O throughputs of four storage structures: HDFS, OrangeFS, Tachyon and two-level storage. We conduct computational experiments to characterize I/O throughput behavior of two-level storage and compare its performance to that of HDFS and OrangeFS, using TeraSort benchmark. Theoretical models and experimental tests both show that the two-level storage system can increase the aggregate I/O throughputs. This work lays a solid foundation for future work in designing and building HPC systems that can provide a better support on I/O intensive workloads with preserving existing computing resources.

## Categories and Subject Descriptors
D.4.3 [**Operating Systems**]: File Systems Management—Distributed file systems

## 1. INTRODUCTION
HPC clusters provide a cost effective computing infrastructure by integrating a large number of commodity computing devices and separate data storage systems. HPC clusters are widely adopted by both industry and academia [3, 4, 6] to support compute-intensive applications. Many software architectures have also been developed to support computing on HPC clusters [2, 22].

Recently, data-intensive computing tasks are growing very fast [9, 10, 15, 36]. To support data-intensive workloads, new software frameworks including distributed data systems (e.g. HDFS [25] and its alternatives [29], LinkedIn Espresso), cluster resource management systems (e.g. YARN in Yahoo, Mesos in Twitter, Helix in LinkedIn, Corona in Facebook), parallel programming models (MapReduce, Tez, Spark, Flink, etc), NoSQL databases and SQL interfaces (HBase, Hive, BlinkDB, and more). large-scale graph and machine learning frameworks (Giraph, GraphX, Gelly, Mahout, MLib, FlinkML), and much more [5, 30], have been developed.

Data-intensive computing has also increasingly become one of the major workloads on HPC clusters [7, 8, 22, 34], and presents a new challenge for traditional HPC architectures [22]. Past investigations have attempted to integrate data-intensive frameworks into HPC cluster [21]. The software frameworks developed for data-intensive computing are usually designed for distributed architectures that differ from the architectures of traditional HPC clusters (see Background section for more details). One important issue in those integrations is how to store the data. Previous studies either used parallel storage on HPC [14, 20, 24, 29, 33] or deployed distributed file systems for data-intensive computing, such as Hadoop distributed file system (HDFS), on compute nodes [11, 26]. Using parallel storage on HPC provides high storage capacity with low cost data fault tolerance, but encounters scalability issues limited by network and aggregate I/O bandwidth of storage nodes. On the other hand, deploying data-intensive file system on HPC compute nodes delivers high aggregate I/O throughput, but suffers the high cost for data fault tolerance and low data storage capacity.

In this work, we explore a new approach to integrate data-intensive software framework with HPC cluster. We develop a new two-level storage system by integrating an in-memory file system, Tachyon [12], with a parallel file system, OrangeFS [17]. Both theoretical modeling and experimental tests show that the two-level storage system can increase the aggregate I/O throughputs. The two-level storage system also maintains the low cost on data fault tolerance and high storage capacity. We expect that the two-level storage approach will provide better performance for big data analytics on HPC clusters.

## 2. BACKGROUND
### 2.1 HPC and Hadoop Architectures
The architecture of distributed systems has a significant impact on the performance of data-intensive computing. In HPC architecture, data nodes are separated from compute nodes and are connected via high-performance network. HPC clusters typically provide two types of data storage services: persistent global-shared parallel file system on data nodes and temporal local file system on compute nodes. Typically, the storage capacity of local file system is small and HPC depends on the global-shared persistent storage for large capacity. During computing, data is transferred from global-shared storage on data nodes to local file systems on compute nodes where computational tasks are executed. Data hosted on local storage devices is ephemeral and purged when jobs complete or resource limit is reached. The separation of compute nodes from data nodes provides easy data sharing, but fails to exploit spatial locality of data, which is critical for achieving scalable data-intensive computing.

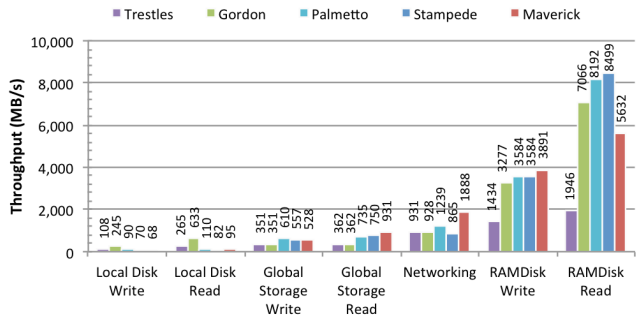

**Figure 1. I/O throughputs of a single compute node on natinal HPC clusters.**

On the other hand, in Hadoop architecture, the compute node and the data node are co-located on the same physical machine. The local storage device on each node is also used as part of the primary persistent data storage. The computational task is scheduled to the physical machine where the required data is stored in order to achieve maximum data locality. The Hadoop system employs the write-once-read-many access model. One of important characteristics of Hadoop system is that its I/O throughput can be accumulated over nodes. Thus, the Hadoop system can achieve higher aggregate I/O throughputs with more nodes.

## 2.2 Characterizing I/O Performance of HPC Clusters

In data-intensive computing, workloads are usually I/O bound. In order to understand the I/O performance of HPC clusters, we measure the throughputs of different I/O operations on both local and global storage systems of four national HPC clusters as well as the Palmetto cluster in Clemson University (Table 1).

HPC systems often use a single hard disk drive (HDD) as the local storage and employ a dedicated parallel file system, such as Lustre or OrangeFS, as global storage. The nodes are connected by either 10 Gbit/sec Ethernet or 20/40 Gbit/sec Infiniband. The memory module on each node is DDR2/DDR3. As shown in Table 1, the size of memory on each node is comparable to the size of its local storage.

We perform a large sequential read and write on storage systems and RAMDisk using Linux built-in tool "*dd*" with a single process. To get actual on-disk throughput, we use *direct I/O* for data access to avoid the buffer cache. The test program writes and reads 16 consecutive files to measure I/O throughputs on local and global storages. Each of the files is setup to 1 GB. However, we use a single 10 GB file on RAMdisk in order to saturate the high bandwidth on the memory device. Each measurement is repeated five times on three different compute nodes. As shown in Figure 1, the average read throughput of a single thread from DRAM is about 10 times larger than those from global storage, which is then 2.65 times larger than the read throughput from local storage. The write throughput with a single thread to DRAM is 6.57 times larger than those to global storage, which is 4 times larger than the write throughput to local storage. The network throughput is measured by using "*Iperf*" based-on IPoIB link-layer. We get a reduced TCP throughput on high-performance Infiniband network due to the low MTU value (2044) hard coded on HPC compute nodes. Higher MTU value should be able to deliver much higher throughput.

**Table 1. Compute Node Storage Space on National HPC Clusters.**

| HPC | Disk (GB) | DRAM (GB) | PFS (GB) | CPU (Core) |
|---|---|---|---|---|
| Stampede | 80 | 32 | $14\times10^6$ | 16 |
| Maverick | 240 | 256 | $20\times10^6$ | 20 |
| Gordon | 280 | 64 | $1.6\times10^6$ | 16 |
| Trestles | 50 | 64 | $1.4\times10^6$ | 32 |
| Palmetto | 900 | 128 | $0.2\times10^6$ | 20 |
| Avg. | 310 | 109 | $7.4\times10^6$ | 21 |

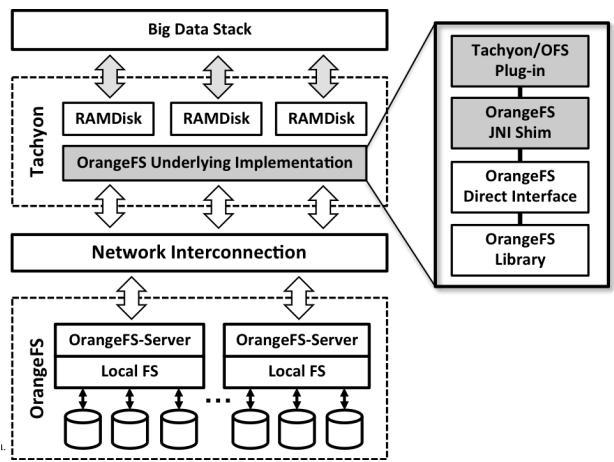

**Figure 2. System architecture with two-level storage system.**

## 3. DESIGN AND IMPLEMENTATION OF TWO-LEVEL STORAGE SYSTEM

Due to architectural difference on data storage, the adoption of data-intensive computing framework, such as Hadoop, on HPC environment has faced significant performance issues. As the capacity of local storage of HPC is small, if we deploy HDFS over local storage system, the size of data that Hadoop can process will be limited. Moreover, if Hadoop use the global parallel file system on HPC as storage, the averaged I/O throughput received by each compute node decreases as the number of compute nodes increases, which affects the scalability of Hadoop system.

To support scalable data-intensive computing on HPC infrastructure using Hadoop, we propose a two-level storage system. We combine an in-memory file system on the compute nodes and a parallel file system on data nodes. As the compute nodes in HPC clusters are often equipped with large memory, the in-memory file system can have a storage capacity comparable to local storage-based HDFS. Moreover, the I/O throughputs of in-memory file system are much larger than those of local disk (Figure 1). We expect that the two-level storage system can improve the aggregate I/O throughputs. Meanwhile, the parallel file system provides the data-fault tolerance and large storage capacity. Then, the two-level storage takes advantages of both in-memory file system and parallel file system.

We implement a prototype of the two-level storage system by integrating Tachyon-0.6.0, an in-memory file system, with the OrangeFS-2.9.0, a parallel file system. Figure 2 shows the architecture of our two-level file storage system. The Tachyon is implemented in Java and the OrangeFS is implemented in C. We introduce two components to tightly integrate Tachyon with OrangeFS (shadow parts in Figure 2).

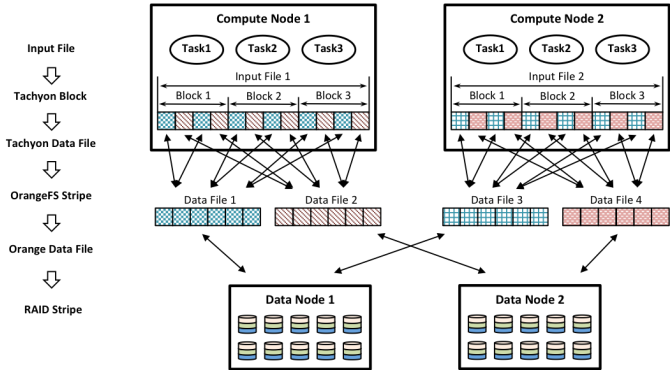

**Figure 3. Input file partition and checkpointing data file striping on two-level storage system.**

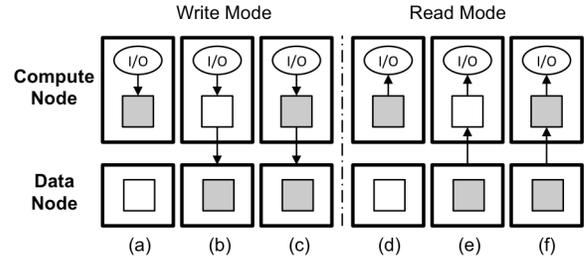

**Figure 4. I/O operation modes of two-level storage.**

- **Tachyon-OFS Plug-in**: a Java plug-in that provides the interface to transform the functionalities of Tachyon in-memory file system to the functionalities of OrangeFS parallel file system. The plug-in also provides *hints* with storage layout support to allow deep tuning between the two file systems.
- **OrangeFS JNI Shim**: a Java API that forwards all function calls from Tachyon-OFS Plug-in to the OrangeFS Direct Interface. To deliver a high bandwidth, the shim layer uses Java Native Interface (JNI) with Non-blocking I/O (NIO) APIs and optimized buffer size to minimize overheads introduced in JVM.

During the development of prototype, we have also introduced new features for OrangeFS and Tachyon projects [27] and contributed our work back to the open source communities. Related patches have been merged into both of OrangeFS trunk [18] and Tachyon master branch [28].

## 3.1 Data Layout Mapping

In our two-level storage system, OrangeFS and Tachyon have different data layouts. As shown in Figure 3, an input file is transparently stored in Tachyon as a set of fixed size logical blocks. The block size controls data-parallel granularity and can be predefined in configuration. In contrast, the data file is stored in OrangeFS as stripes. Each OrangeFS data file is then striped in disk-level, which is usually performed by hardware RAID (redundant array of independent disks) built in each data node. Data fault tolerance of the two-level storage system is ensured by the low-level erasure coding inside each data node.

When data is written from Tachyon to OrangeFS or read from OrangeFS to Tachyon, the data is mapped to the layout of the target file system. This mapping between Tachyon and OrangeFS data layouts can impact the load balance among data nodes and the aggregate I/O throughputs of compute nodes. In order to achieve optimal performance of our two-level storage system, several parameters, such as block size of Tachyon and stripe size of OrangeFS, need to be tuned. The parameters of Tachyon are specified in configuration files and read during the start of Tachyon. The parameters of OrangeFS can be dynamically changed through hints implemented in our Plug-in.

## 3.2 I/O Modes of Two-level Storage System

Currently, we have implemented synchronous I/O on the prototype two-level storage system. The prototype provides three write modes and three read modes (Figure 4). The three write modes are: data is stored only in Tachyon; data bypasses Tachyon and is written to OrangeFS; and data is synchronously written to OrangeFS when data is created or updated in Tachyon. These three modes are depicted in Figure 4 (a-c) respectively. The read modes are similar and shown in Figure 4 (d-f). Specifically, they are data is read from Tachyon only; data is read from OrangeFS directly without caching in Tachyon; and data is read partially from Tachyon and partially from OrangeFS. The read mode in Figure 4 (f) is the primary usage pattern in data-intensive computing. It improves read performance by caching reusable data and adopting a proper data eviction policy such as LRU/LFU.

Reading data from remote data nodes, especially from overloaded data nodes, is very expensive. To minimize I/O congestion and contention, we apply priority-based read policy on two adjustable I/O buffer caches, one between application and Tachyon and another one between Tachyon and OrangeFS. The read I/O request is always sent to next available storage device with shortest distance (where the targeted data is hosted). Since Hadoop schedules computing tasks based on data locality, most of the computing tasks first fetch the input data from local Tachyon file system. If the data cannot be found in Tachyon, the read request is forwarded to load the check pointed block from OrangeFS persistent storage layer. To get an optimized I/O throughput and latency, we choose 1 MB I/O buffer to request Tachyon data and 4 MB I/O buffer for loading data from OrangeFS file system.

## 4. MODELING I/O THROUGHPUTS OF DIFFERENT STORAGES

We consider a HPC system consisting of $N$ compute nodes and $M$ data nodes. We make the following simplifying assumptions in our modeling effort:

- All nodes have identical hardware configurations and are connected via non-blocking switches.
- The centralized switch and the bisection bandwidth of network are able to provide a non-blocking backplane throughput $\Phi$ and each node is attached by a full-duplex network interface with a bandwidth throughput $\rho$.
- There is no network-level interference, such as TCP congestion, and Incast/Outcast.

**Table 2. The List of Notation Abbreviations.**

| Symbol | Significance |
|---|---|
| $D$ | Data size |
| $N$ | Number of compute nodes |
| $M$ | Number of data nodes |
| $f$ | The ratio of the size of data in Tachyon over the total size of data |
| $\Phi$ | Bandwidth of switch backplane, bisection bandwidth of network (MB/s) |
| $\rho$ | Bandwidth of network interface of compute and data nodes (MB/s) |
| $\mu$ | I/O throughput of local hard drives on compute nodes (MB/s) |
| $\mu'$ | I/O throughput of local hard drives on data nodes (MB/s) |
| $\nu$ | I/O throughput of local memory (MB/s) |
| $q$ | Average I/O throughput received on compute nodes (MB/s) |

We deploy Hadoop on $N$ compute nodes of HPC for data-intensive computing. All computational tasks of interests are I/O bounded and evenly distributed on each compute node without data skew. Hadoop can use four different types of storages: HDFS, OrangeFS, Tachyon and the two-level storage system. We model the I/O throughputs of each compute node with each of these four different types of storages when it reads/writes a fixed size $D$ of data from/to storage. The notations used in the models are listed in Table 2.

## 4.1 I/O Modeling of HDFS
With HDFS, Hadoop reads mainly from the local hard drives. In this case, the read throughput $q_{read}^{HDFS}$ of each compute node is determined by the I/O throughput to local hard drive, $\mu$. If the data is not available on local hard drive, Hadoop reads from other nodes through network. In this case, the $q_{read}^{HDFS}$ is determined by minimum throughput of following three factors: bandwidth throughput of network interface of each node, $\rho$; shared backplane throughput, $\Phi/N$; and I/O throughput to local hard drive, $\mu$. Then, the read throughput of each node, $q_{read}^{HDFS}$, is

$$q_{read}^{HDFS} = \begin{cases} \mu, & local\ access \\ min\left(\rho, \frac{1}{N}\Phi, \mu\right), & remote\ access \end{cases} \quad (1)$$

To maintain the fault-tolerance of data, by default, Hadoop synchronously writes one copy of data to local hard drive and two mirrored copies of data to other two nodes by streaming through network. Thus, the write throughput of each node, $q_{write}^{HDFS}$, is also determined by the minimum throughput of three factors: bandwidth throughput of network interface of each node, shared backplane throughput and the I/O throughput of local hard drive. Considering the whole cluster, all nodes write three copies of data to local storage. Then, the maximum write throughput of each node to local hard drive is $\frac{1}{3}\mu$. Each node writes two copies of data to network. Thus, throughput of network interface of each node is limited by $\frac{1}{2}\rho$ and average throughput of bisection backplane is bounded by $\frac{1}{2N}\Phi$. Thus, the write throughput of each node, $q_{write}^{HDFS}$, can be estimated as

$$q_{write}^{HDFS} = min\left(\frac{1}{2}\rho, \frac{1}{2N}\Phi, \frac{1}{3}\mu\right) \quad (2)$$

## 4.2 I/O Modeling of OrangeFS
With OrangeFS as storage for Hadoop, the $N$ compute nodes read/write data from $M$ data nodes. All read and write traffics must pass through the network. Thus, both write and read throughputs are determined by the throughputs of following four resources: (1) the bandwidth of network interface of a compute node, $\rho$. (2) the shared throughput on bisection backplane; since all nodes share the bandwidth of the switch backplane, the average throughput received by each compute node is $\frac{1}{N}\Phi$ assuming $N > M$. (3) the shared throughput of network interface of data nodes; aggregate throughput of network interface of $M$ data nodes is $M \times \rho$, shared by $N$ compute nodes. Thus, the average network interface throughput of data nodes that each compute node receives is $\frac{M}{N}\rho$. (4) the shared I/O throughput to local hard drive in data nodes; aggregate local hard drive I/O throughput of $M$ data nodes is $M \times \mu'$, shared by $N$ compute nodes. Thus, the average I/O throughput to local hard drive on data nodes is $\frac{M}{N}\mu'$. Together, the read throughput, $q_{read}^{OFS}$, and write throughput, $q_{write}^{OFS}$, of each compute node are

$$q_{write}^{OFS} = q_{read}^{OFS} = min\left(\rho, \frac{1}{N}\Phi, \frac{M}{N}\rho, \frac{M}{N}\mu'\right) \quad (3)$$

## 4.3 I/O Modeling of Tachyon
The Tachyon system has similar architecture as HDFS except that: (1) it uses the RAM, rather than local hard drive, to store data; (2) it uses linage-based recovery, rather than data replication, to achieve data fault-tolerance, which can significantly improve the write throughput.

With Tachyon as storage, Hadoop reads from RAM of each compute node. In this case the read throughput $q_{read}^{Tachyon}$ of each compute node is determined by the I/O throughputs to RAM, $\nu$. If the data is not available from local RAM, Hadoop reads from RAM of other nodes in the network. In this case, the $q_{read}^{Tachyon}$ is determined by the minimum of three throughputs: network interface bandwidth, $\rho$; shared backplane bandwidth, $\Phi/N$; and the I/O throughput $\nu$ to RAM. Namely, the read throughput of each node, $q_{read}^{Tachyon}$, is

$$q_{read}^{Tachyon} = \begin{cases} \nu, & local \\ min\left(\rho, \frac{1}{N}\Phi, \nu\right), & remote \end{cases} \quad (4)$$

With Tachyon as storage, Hadoop will write the data to RAM of each compute node. Then, the write throughput of each compute node, $q_{write}^{Tachyon}$ is just limited by the throughput to memory:

$$q_{write}^{Tachyon} = \nu \quad (5)$$

## 4.4 I/O Modeling of The Proposed Two-level Storage
To simplify our analysis, we assume Hadoop uses the third write mode in Figure 4 (c) and the third read mode in Figure 4 (f) of the two-level storage system. Thus, the Hadoop reads/writes data from/to both Tachyon and OrangeFS. For third write mode, the data is synchronously written to Tachyon and OrangeFS at the same time. As the write throughput to Tachyon is much higher than those to OrangeFS, the write throughput of each compute node on two-level storage, $q_{write}^{TLS}$, is bounded by the write throughput to OrangeFS:

$$q_{write}^{TLS} = min(q_{write}^{Tachyon}, q_{write}^{OFS}) = q_{write}^{OFS} \quad (6)$$

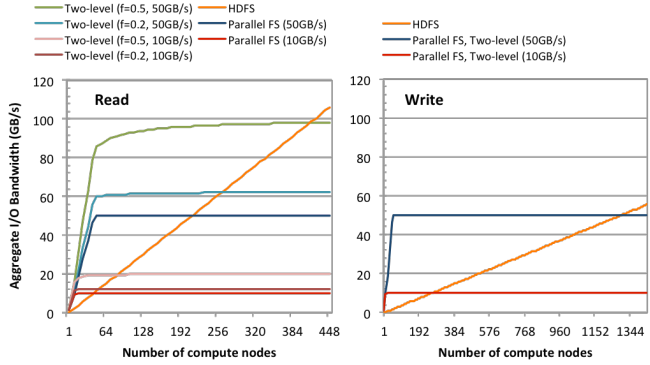

**Figure 5. Aggregate read throughput (left) and write throughput (right) of HDFS, parallel file system and two-level storage.**

Let $f$ be the ratio of the size of data in Tachyon over the total size of data, $D$. Then, the size of data in Tachyon is $f \times D$ and the size of data in OrangeFS is $(1-f) \times D$. The Hadoop will read $f \times D$ data from Tachyon with throughput $v$ (Tachyon in the two-level storage do not read the data from other compute nodes) and $(1-f) \times D$ data from OrangeFS with throughput $q_{read}^{OFS}$. Together, the read throughput of each compute node is

$$q_{read}^{TLS} = 1/(\frac{f}{v} + \frac{1-f}{q_{read}^{OFS}}) \qquad (7)$$

If $f = 1$, all data is read from Tachyon only and if $f = 0$, all data is read from OrangeFS only. The higher the value of $f$, the higher read throughput the two-level system can provide.

## 4.5 Comparing Aggregate I/O Throughputs of Different Storages

The aggregate read/write throughputs of HDFS can linearly scale up with the number of compute nodes. On the other hand, the aggregate read/write throughputs of parallel file systems (such as OrangeFS) and the two-level storage are bounded by the network bandwidth and aggregate throughput of local disks on data nodes. To understand the aggregate I/O throughputs of parallel file systems and two-level storage comparing to those of HDFS, we have a case study using the average I/O throughputs of HPC clusters (Figure 1). The network bandwidth is set to 1,170 MB/s per node. The local disk read throughput is 237 MB/s and the local disk write throughput is 116 MB/s. The local memory throughput is 6,267 MB/s. We have tested two parallel file system aggregate throughputs: 10 GB/s and 50 GB/s. We assume the HDFS is deployed on single hard disk of compute nodes of HPC. We don't consider the storage capacity that systems can support, but focus only on the throughput study.

We calculate the aggregate read/write throughputs of three storages. As shown in Figure 5, at 10 GB/s aggregate bandwidth of parallel file system, HDFS needs only 43 nodes to achieve a higher aggregate read bandwidth than that parallel file system and needs 53 nodes ($f = 0.2$) and 83 nodes ($f = 0.5$) to achieve higher read aggregate bandwidths than that the two-level storage has. At 50 GB/s aggregate bandwidth of parallel file system, the HDFS needs 211 nodes to have a higher aggregate read bandwidth than that parallel file system has and needs 262 nodes ($f = 0.2$) and 414 nodes ($f = 0.5$) to have higher aggregate read bandwidths than that the two-level storage has. Our results show that the two-level storage has increased the aggregate read bandwidth by about 25% at $f = 0.2$ (from 10 GB/s to 12.5 GB/s or from 50 GB/s to 62GB/s) and about 95% at $f = 0.5$ (from 10

**Table 3. Hardware Configurations of Selected Nodes on Palmetto Cluster.**

| CPU | Intel Xeon E5-2670 v2 20×2.50 GHz |
|---|---|
| HDD | 1 TB 7200RPM SATA |
| RAID | 12 TB LSI Logic MegaRAID SAS |
| RAM | 128 GB DDR3-1600 |
| Network | Intel 10 Gigabit Ethernet |
| Switch | Brocade MLXe-32 with 6.4 Tbps backplane |

GB/s to 19.6 GB/s or from 50 GB/s to 98 GB/s). Thus, using the two-level storage can increase the number of compute nodes to deploy Hadoop without sacrificing read performance.

Meanwhile, at 10 GB/s aggregate bandwidth of parallel file system, HDFS needs 259 nodes to have higher aggregate write bandwidth than those parallel file system and the two-level storage have, and at 50 GB/s aggregate bandwidth of parallel file system, HDFS needs 1,294 nodes to have higher aggregate write bandwidth than those parallel file system and the two-level storage have. The write throughput of HDFS is much smaller than read throughput because Hadoop needs to write two copies of data through network. Thus, write throughput is usually not the constraint to use Hadoop on HPC with parallel or two-level storages.

## 5. EXPERIMENTAL EVALUATION

In this section, we evaluate our two-level storage system using two experiments. We first characterize I/O throughput behavior of the two-level storage. Then, we compare performance of CPU, disk and network I/O utilizations of each compute node and the performance of disk and network I/O utilizations of each data node using TeraSort benchmark program when Hadoop is deployed on HDFS, OrangeFS and the two-level storage, respectively.

### 5.1 Experimental Setup

All experiments are performed on Palmetto HPC cluster hosted at Clemson University. We select nodes with the same hardware configuration (Table 3) for our experiments. Each compute node is attached with a single SATA hard disk, and each data node is attached with 12 TB disk array. Although we cannot control the bandwidth of switch backplane, the backplane bandwidth is much higher than the network interface bandwidth in our experiments and is not the bottleneck resource in our experiments.

For the first experiment, we use Tachyon built-in performance evaluation program as the benchmark tool to measure the average read throughput received from two-level storage under a range of *data sizes* with different *skip sizes*. In the experiment, we conduct our measurements between one compute node and one data node. We allocate 16 GB for Tachyon storage space on compute node and the data node has a 12 TB OrangeFS file system. The skip size is defined as a fragment of data skipped per MB access. Since OrangeFS has much higher access latency than Tachyon has, a large skip size has larger impact on the I/O throughput for OrangeFS than for Tachyon. The data size is varied from 1 GB to 256 GB. For each data size, we test a range of skip sizes from 0 KB to 64 MB.

For the second experiment, we run the Terasort benchmark on a 17-node Hadoop cluster with 2-node OrangeFS as back-end storage system. In Hadoop cluster, we use one machine as the head node to host YARN's ResourceManager (RM) and

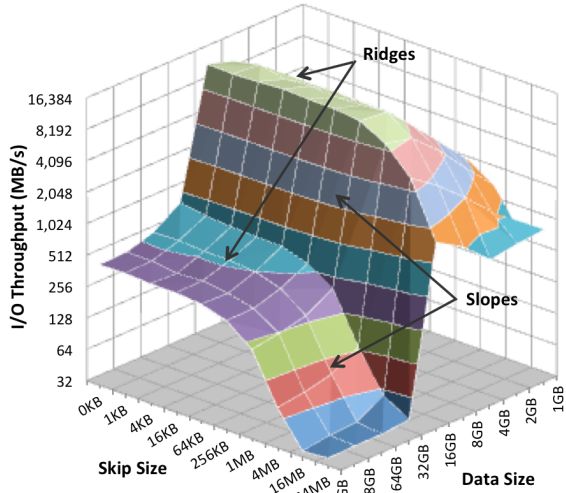

**Figure 6. The storage mountain of two-level storage system.**

Tachyon's Master service. Then, we have 16 compute nodes in Hadoop cluster. On each compute node, we assign 16 containers to occupy 16 CPU slots and leave the rest of 4 CPU slots to handle extra system overhead. Thus, we can run 256 Mappers or Reducers and the workload can achieve full system utilization if CPU utilization reaches 80%.

The capacity of Tachyon storage on each compute node is 32 GB. Then, the total capacity of Tachyon is 512 GB. The Tachyon block size is set to 512 MB. Each block is striped into 8 chunks with a strip size of 64 MB that are evenly distributed across 2 data nodes with round-robin fashion. To get an optimized I/O throughput on two-level storage, we use 1 MB request size for MapReduce applications and 4 MB I/O buffer between Tachyon and OrangeFS. The request size and buffer size are selected by performing a series of I/O throughput measurements.

Before each test, we empty OS page caches to measure actual I/O costs. The concurrent write and read throughputs on local disk for each of compute nodes are about 60 MB/s. Concurrent write throughputs on RAID for each of data nodes is about 200MB/s, and read is 400 MB/s.

## 5.2 Characterizing The I/O Performance of Two-level Storage

As illustrated in Figure 6, we generate a two dimensional function of read throughput versus data size and skip size. This function is similar to the memory mountain that characterizes the capabilities of memory system. We call this function the *storage mountain* of two-level storage system. The storage mountain reveals the performance characteristics of our prototype two-level storage system. There are two ridges on the storage mountain. The high ridge corresponds to throughput of Tachyon and the low ridge reflects the throughput of OrangeFS. There is a slope between the two ridges when the data size is larger than 16 GB, which is the size of Tachyon storage. During our experiments, the I/O buffer size between applications and Tachyon is set to 1 MB. There are slopes on both ridges when skip size is larger than 1 MB. The read throughputs decrease when the data size is small. This is because the extra overhead cost, such as scheduling cost, data sterilization, become noticeable when the I/O cost of small data is low.

The storage mountain shows that the performance of the two-level storage is affected by multiple factors, such as data size and buffer size. Since the ridge of Tachyon is much higher than that of OrangeFS, we need to keep frequently used data in Tachyon to achieve better performance.

## 5.3 Evaluating Performance Using TeraSort

In this experiment, we profile the detailed performance metrics with the TeraSort benchmark workload. The TeraSort benchmark has three stages: *TeraGen* stage generates and writes input data to storage; *TeraSort* stage loads input data, sorts and writes output data to storage; and *TeraValidate* stage reads and validates the sorted output data. Since the TeraSort stage reads once and writes once and is an I/O bounded task, we use this stage to evaluate I/O performance of three storages: HDFS, OrangeFS and two-level storage.

We first run the TeraGen stage using a Map-only job to generate 256 GB data and store to three storages: HDFS, OrangeFS and two-level storage (one copy in Tachyon and one copy in OrangeFS). We then run the TeraSort stage using one Map/Reduce cycle. Mapper reads the data from storage and Reducer writes the sorted data back to storage. We profile the performance of CPU, disk and network I/O utilizations of each compute node and the performance of disk and network I/O

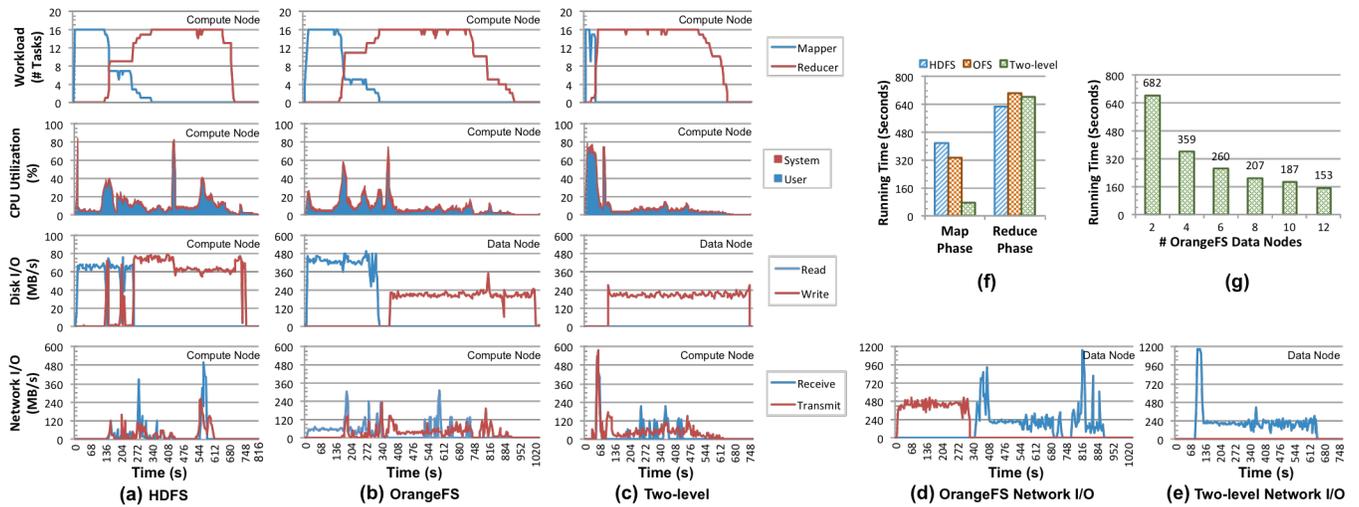

**Figure 7. Performance profiling metrics for TeraSort benchmark suit on three storage systems.**

utilizations of each data node (Figure 7 (a-e)).

With HDFS, the Mapper reads from and the Reducer writes to local disks on compute nodes. With OrangeFS, the Mapper reads from and the Reducer writes to OrangeFS on data nodes. With the two-level storage, the Mapper reads from Tachyon (RAM) on compute nodes and the Reducer writes to OrangeFS on data nodes.

Since we can store all data in Tachyon of two-level storage in our experiments, the Mapper can achieve peak read throughput (Tachyon ridge in storage mountain) as shown Figure 7 (f). The Mapper on two-level storage is able to achieve about 5.4× and 4.2× speedup comparing to the Mapper on HDFS and OrangeFS, respectively (Figure 7 (f)). The high read throughput even pushes the Mapper reaching full CPU usage (Figure 7 (c)). Keeping part of data in Tachyon of two-level storage also reduces the network traffic. In our extreme case, there is no network traffic from data nodes for Mappers using two-level storage (Figure 7 (e)).

Writing to OrangeFS through Tachyon can also slightly improve the performance comparing to directly writing to OrangeFS (Figure 7 (b, c, f)). It benefits from unidirectional I/O access from Tachyon to OrangeFS, in which OS page caches of data nodes can fully engage in optimizing write loads. As a comparison, the data nodes are involved for handling both read and write loads when only OrangeFS is used. The Reducer running time on OrangeFS and two-level storage is slightly longer than that using HDFS (Figure 7 (f)) when we use only two data nodes. However, the write throughputs of OrangeFS and two-level storage can be steadily improved by scaling the data node. For example, when a new data node to our testing system is added, roughly an extra 200 MB/s concurrent write throughput can be achieved. Running time of TeraSort reduce phase decreases by 1.9× and 4.5× when the number of data nodes increases from 2 to 4 and 12 respectively are (Figure 7 (g)). In all tests, performance is bounded by either aggregate disk throughput or CPU FLOPs of compute nodes, rather than networking bandwidth. As shown in Figure 7, the network throughput never reaches the limitation.

The data used in our current experiments is relatively small and can be completely stored in Tachyon of two-level storage. If we have large size of data that is stored in both Tachyon and OrangeFS of two-level storage, the performance of TeraSort using two-level storage will be degraded gracefully. However, according to our theoretical analysis (Figure 5), we still expect that the two-level storage could always have better performance than OrangeFS. Meanwhile, the two-level storage is able to delivery higher I/O throughputs and much larger storage capacities than HDFS under limited numbers of compute nodes.

## 6. RELATED WORK

There are three major research directions to integrate Hadoop with HPC infrastructure. Previous work has explored directly deploying Hadoop atop of existing parallel file systems, such as GPFS [1], Ceph [14], Lustre [23]. These efforts mainly focus on showing the performance enhancement by exposing suitable mapping between parallel file systems and Hadoop, such as increasing the size of stripe unit, using different layout distribution, and applying optimal data prefetching. However, the performance of data-intensive workload is still tightly coupled with available I/O bandwidth of parallel file systems.

Instead of using dedicated data servers, some of previous work deploys Hadoop on compute nodes only. Tantisiriroj et al. [29] explore the I/O performance benefit by migrating data server to compute nodes with emulated HDFS-style data layout, replication and consistency semantics. In their experiments, the performance of PVFS (v2.8.2) is very close and even higher than that of HDFS (v0.20.1) on 51-node OpenCloud cluster when using optimized I/O buffer size, data mapping and layout. Other researches have deployed parallel file system, Gfarm and GlusterFS [16] as well as QFS [20], on compute nodes in their production cluster. However, the capacity, performance and non-consistence of local disk in traditional HPC limit the usability of deploying Hadoop on compute nodes.

Third solution deploys Hadoop on data nodes. Xu et al. [35] have studied performance enhancement by employing MapReduce on the storage sever of HPC. This deployment solution can access data on persistent storage natively. It works on a small size of workloads but it could have scalability issue when the job has mixed CPU and data-intensive workloads. Because data nodes on HPC usually are equipped with relatively slow and limited computing units. This is especially true for many of CPU-bound data analysis workloads [19].

One project [32], is similar in spirit to our project, but from a different direction. It uses two optimized schedule techniques (Enhanced Load Balancer and the Congestion-Aware Task Dispatching) to improve the I/O performance of local disk. Our solution is focusing on integration of two storage systems.

Wang et al. [31] have also utilized memory to increase I/O performance of parallel file system. They introduce a dedicated buffer layer deployed at the front-end of data nodes of HPC to buffer the burst I/O. In our system, we use the memory of compute nodes as part of storage.

Recently, the search engine, Baidu, reported use of Tachyon as a transparent layer for data exchange between Baidu file system (BFS) hosted in data centers in China and those in USA research center [13]. Depending on the workload type, overall improvement was 30 to 60 time speedups.

## 7. CONCLUSSION AND DISCUSSION

In this paper, we develop a prototype of two-level storage by integrating the in-memory file system, Tachyon, and the parallel file system, OrangeFS. In two-level storage, Tachyon is deployed on compute nodes and OrangeFS is deployed on data nodes. Tachyon provides a temporal locality of data that is not needed to retrieve from data nodes through network. Our theoretical modeling and experimental evaluation show that the current version of two-level storage can increase read throughput. Since write throughput is usually not a bottleneck for running Hadoop on HPC, higher read throughput of the two-level storage will scale up with the number of compute nodes for Hadoop.

Although running Hadoop on Tachyon alone can also take advantage of high I/O throughput and data locality, it has two issues. First, the capacity of Tachyon is limited comparing to large storage capacity on data nodes. Second, Tachyon uses lineage to recover data when there is a fault. This recovery incurs computing cost. In our two-level storage, local data always has a copy in OrangeFS; thus, OrangeFS provides fault-tolerance for Tachyon.

Public HPC clusters are usually shared by a lot of users. Each user is usually allocated a limited number of compute nodes. The two-level storage can provide higher read and write throughput with limited number of compute nodes. Thus, running Hadoop with the two-level storage may provide a better performance solution for big data analytics on traditional HPC infrastructures.

## 8. ACKNOWLEDGMENTS
We thank the supporting from OrangeFS community led by Dr. Walter B. Ligon. We are thankful to Haoyuan Li from Tachyon Nexus, Inc., and Zhao Zhang from UC Berkeley, AMPLab and BIDS for providing helpful feedbacks. HPC resources used in this research are supported by Omnibond Systems, LLC and the Clemson Computing and Information Technology (CCIT). This work was partially supported by the NSF under Grant No. CCF-1551511.